\documentclass[12pt]{article}
\usepackage[dvips]{epsfig}
\usepackage{harvard}
\citationmode{abbr}
\usepackage{amsmath, amsthm, amssymb, graphicx}
\usepackage{enumerate}
\usepackage{float}
\usepackage{bm}
\usepackage{mathrsfs}

\renewcommand*\&{and}

\begin{document}

\title{NBLDA: Negative Binomial Linear Discriminant Analysis for RNA-Seq Data}
\author{Kai Dong$^1$, Hongyu Zhao$^3$, Xiang Wan$^{2,}$\thanks{Co-corresponding author. Email: xwan@comp.hkbu.edu.hk}
~and Tiejun Tong$^{1,}$\thanks{Co-corresponding author. Email: tongt@hkbu.edu.hk} \\
\\
{\small $^1$Department of Mathematics, Hong Kong Baptist University, Hong Kong} \\
{\small $^2$Department of Computer Science, Hong Kong Baptist University, Hong Kong} \\
{\small $^3$Department of Biostatistics, Yale University, New Haven, CT 06510, USA}
}

\date{\today}
\maketitle

\baselineskip 24pt

\begin{abstract}

RNA-sequencing (RNA-Seq) has become a powerful technology to characterize gene expression profiles because it is more accurate and comprehensive than microarrays.
Although statistical methods that have been developed for microarray data can be applied to RNA-Seq data, they are not ideal due to the  discrete nature of RNA-Seq data.
The Poisson distribution and negative binomial distribution are commonly used to model count data.
Recently, \citeasnoun{witten2011classification} proposed a Poisson
linear discriminant analysis for RNA-Seq data.
The Poisson assumption may not be as appropriate as negative binomial distribution when biological replicates are available and in the presence of overdispersion (i.e., when the variance is larger than the mean).
However, it is more complicated to model negative binomial variables because they involve a dispersion parameter that needs to be estimated.

In this paper, we propose a negative binomial linear discriminant analysis for RNA-Seq data.
By Bayes' rule, we construct the classifier by fitting a negative binomial model,
and propose some plug-in rules to estimate the unknown parameters in the classifier.
The relationship between the negative binomial classifier and the Poisson classifier is explored, with a numerical investigation of the impact of dispersion on the discriminant score. Simulation results show the superiority of our proposed method. We also analyze four real RNA-Seq data sets to demonstrate the advantage of our method in real-world applications.

\end{abstract}

\section{Introduction}
RNA-sequencing (RNA-Seq) is a revolutionary technology that uses the capabilities of next-generation sequencing to quantify gene  expression levels \cite{mardis2008next,wang2009rna,morozova2009applications}.
Compared to microarray technology, RNA-Seq has many advantages including the detection of novel transcripts, low background signal, and the increased specificity and sensitivity.
Due to reduced sequencing cost, RNA-Seq has been widely used in biomedical research in recent years \citeasnoun{lorenz2014using}.
In real-world applications, the gene expression profile of biopsy or serum sample from an individual can be used to test whether this individual has a disease and/or a specific type of disease, which is essentially a classification problem.
Different from the microarray technology that measures the level of gene expression on a continuous scale,
RNA-Seq counts the number of reads that are mapped to one gene and measures the level of gene expression with nonnegative integers. As a result, popular tools that assume a Gaussian distribution in microrray data analysis, such as linear discriminant analysis, may not perform as well as those methods that adopt appropriate discrete distributions for RNA-Seq data.

For RNA-Seq data, the Poisson distribution and negative binomial distribution are two common distributions considered in the expression detection and classification. Many methods have been proposed to detect differentially expressed genes, including edgeR \cite{robinson2008small},\citeasnoun{robinson2010edger},
DESeq \citeasnoun{anders2010differential}, baySeq \citeasnoun{hardcastle2010bayseq}, BBSeq \citeasnoun{zhou2011powerful}, SAMseq \citeasnoun{li2013finding},
DSS \citeasnoun{wu2013new}, AMAP \citeasnoun{si2013optimal}, sSeq \citeasnoun{yu2013shrinkage}, and LFCseq \citeasnoun{lin2014lfcseq}. However, there is less progress on the classification using RNA-Seq data until recently. \citeasnoun{witten2011classification}  proposed a Poisson linear discriminant analysis (PLDA) which assumes that RNA-Seq data follow the Poisson distribution.
\citeasnoun{tan2014classification} further discussed many methods, such as logistic regression and partial least squares,
and showed that PLDA is a comparable method. The Poisson distribution is suitable for modeling RNA-Seq data when biological replicates are not available.
However, if biological replicates are available, the Poisson distribution may not be a proper choice owing to the overdispersion issue, where the variances of such data are likely to exceed their means \cite{anders2010differential,si2013optimal}. The overdispersion issue can have a significant effect on classification accuracies. In real-world applications, biological replicates can provide more convincing results than technical replicates.
Therefore, it is necessary to look for some solutions to take the overdispersion issue into consideration.

We note that \citeasnoun{witten2011classification} has considered this problem and pointed out that the classification accuracy can be further improved for overdispersed data by extending the Poisson model to the negative binomial model.
However, to construct an appropriate negative binomial classifier for practical use, two major issues remain to be solved.
The first issue is that the probability density function (pdf) of the negative binomial distribution is more complicated than that of the Poisson distribution, which gives rise to a more complicated classifier.
The second issue is that the negative binomial distribution contains a dispersion parameter,
which controls how much its variance exceeds its mean.
To construct the classifier using the negative binomial model, we need to estimate the dispersion parameter.
To avoid fitting the complicated negative binomial model, \citeasnoun{witten2011classification} proposed a transformation method for the overdispersed data and found that this method works well if the overdispersion is mild.

In light of the importance of the dispersion in modelling RNA-Seq data with the negative binomial distribution, some dispersion estimation methods have been proposed recently in the literature. For example, \citeasnoun{wu2013new} proposed a dispersion estimator using empirical Bayes method and applied it to find differentially expressed genes.
\citeasnoun{yu2013shrinkage} proposed a shrinkage estimator of dispersion which shrinks the estimates obtained by the method of moments towards a target value,
and also applied it to detect differentially expressed genes. These new methods on estimating the dispersion parameter make it possible to construct a negative binomial classifier to achieve better classification accuracy on RNA-Seq data.

In this paper, we propose a negative binomial linear discriminant analysis (NBLDA) for RNA-Seq data.
The main contributions of this paper are in, but not limited to, the following two aspects:
\begin{enumerate}

\item We extend \citeasnoun{witten2011classification} to build a new classifier based on the negative binomial model. Under the assumption of independent genes, we define the discriminant score by Bayes' rule and propose some plug-in rules for estimating the unknown parameters in the classifier.

\item We further explore the relationship between NBLDA and PLDA.
A numerical comparison is conducted to explore how the dispersion changes the discriminant score.
The comparison results will provide some guidelines for scientists to decide which method should be used in the discriminant analysis of RNA-Seq data.

\end{enumerate}
To demonstrate the performance of our proposed method,  we conduct several simulation studies under different numbers of genes, sample sizes, and proportions of differentially expressed genes. Simulation results show that the proposed NBLDA outperforms existing methods in many settings.
Four real RNA-Seq data sets are also analyzed to demonstrate the advantage of NBLDA.
Specifically, we propose the negative binomial classifier study, the relationship between NBLDA and PLDA, and present the parameter estimation in Section 2.
Simulation studies and real data analysis are conducted in Sections 3 and 4. We conclude the paper with some discussions in Section 5.



%
%

\section{Negative Binomial Linear Discriminant Analysis}

Let $X_{ig}$ denote the numbers of reads mapped to gene $g$ in sample $i$, $i=1,\dots,n$ and $g=1,\dots,G$.
Our goal is to identify which class a new observation belongs to.
\citeasnoun{witten2011classification} proposed a PLDA for classifying RNA-Seq data.
When biological replicates are available, however, overdispersion occurs for RNA-Seq data
and hence the Poisson distribution may no longer be appropriate.
In this section, we propose a new discriminant analysis for RNA-Seq data by assuming that the data follow the negative binomial distribution.

\subsection{Methodology}

Consider the following negative binomial distribution for RNA-Seq data:
\begin{eqnarray}\label{nb}
X_{ig} ~\sim~ {\rm NB}(\mu_{ig},\phi_g),~~~~~\mu_{ig}=s_i \lambda_g,
\end{eqnarray}
where $s_i$ is the size factor which is used to scale gene counts for the $i$th sample due to different sequencing depth,
$\lambda_g$ is the total number of reads per gene, and $\phi_g\geq0$ is the dispersion parameter.
We have $E(X_{ig})=\mu_{ig}$ and ${\rm Var}(X_{ig})=\mu_{ig}+\mu_{ig}^2\phi_g$.
Note that the variance is larger than the mean for the negative binomial distribution.

Let $K$ be the total number of classes and $C_k\in\{1,\dots,n\}$ the indices of samples in class $k$ for $k=1,\dots,K$.
Then the class-specific model for RNA-Seq data is given by
\begin{eqnarray}\label{nbc}
(X_{ig}|y_i=k) ~\sim ~ {\rm NB}(\mu_{ig}d_{kg},\phi_g),
\end{eqnarray}
where $d_{kg}$ represents the differences among $K$ classes, and $y_i=k\in\{1,\dots,K\}$ represents the label of sample $i$.
We also follow the independence assumption in \citeasnoun{witten2011classification} that all genes are independent of each other.
Note that the independence assumption is frequently assumed in microarray data analysis.

Let ${\bf x}^*=(X_1^*,\dots,X_G^*)^T$ be a test sample with $s^*$ the size factor and $y^*$ the class label.
By Bayes' rule, we have
\begin{eqnarray}\label{bay}
P(y^*=k|{\bf x}^*) ~\propto~ f_k({\bf x}^*)\pi_k,
\end{eqnarray}
where $f_k$ is the pdf of the sample in class $k$,
and $\pi_k$ is the prior probability that one sample comes from class $k$.
The pdf of $X_{ig}=x_{ig}$ in model (\ref{nbc}) is
\begin{eqnarray}\label{pdf}
P(X_{ig}=x_{ig}|y_i=k) &=& \frac{\Gamma(x_{ig}+\phi_g^{-1})}{x_{ig}!\Gamma(\phi_g^{-1})}
\left(\frac{s_i \lambda_g d_{kg} \phi_g}{1 + s_i \lambda_g d_{kg} \phi_g}\right)^{x_{ig}} \nonumber \\
    &&\left(\frac{1}{1+s_i\lambda_gd_{kg}\phi_g}\right)^{\phi_g^{-1}}.
\end{eqnarray}
By (\ref{bay}) and (\ref{pdf}), we have the following discriminant score for NBLDA:
\begin{eqnarray}\label{rule.1}
\log P(y^*=k|{\bf x}^*) &=& \sum_{g=1}^G X_g^* \left[ \log d_{kg} - \log(1 + s^*\lambda_gd_{kg}\phi_g)\right]  \nonumber \\
                        && -\sum_{g=1}^G \phi_g^{-1} \log(1+s^*\lambda_gd_{kg}\phi_g) \nonumber \\
                        && + \log\pi_k + C,
\end{eqnarray}
where $C$ is a constant independent of $k$.
We then assign the new observation ${\bf x}^*$ to class $k$ that maximizes the quantity (\ref{rule.1}).
Throughout the paper, we estimate the prior probability $\pi_k$ by $n_k/n$, where $n_k$ is the sample size in class $k$.
For balanced data, the prior probability is simplified as $\pi_k=1/K$ for all $k=1,\dots,K$.
For gene $g$, the total number of reads is $\lambda_g=\sum_{i=1}^n X_{ig}$,
and the class difference $d_{kg}$ can be estimated by $(\sum_{i\in C_k}X_{ig} +1)/(\sum_{i\in C_k}s_i\lambda_g +1)$.
Estimation of the unknown parameters including $s_i$ and $\phi_g$ will be discussed in Section 2.2.

To explore the relationship between the proposed NBLDA and the PLDA in \citeasnoun{witten2011classification},
we assume that $s^*\lambda_gd_{kg}$ are bounded.
When $\phi_g\to 0$, we have $\log(1+s^*\lambda_gd_{kg}\phi_g)\to 0$ and
$\phi_g^{-1} \log(1+s^*\lambda_gd_{kg}\phi_g) = \log(1+s^*\lambda_gd_{kg}\phi_g)^{\phi_g^{-1}} \to s^*\lambda_gd_{kg}$.
Then consequently,
\begin{eqnarray}\label{plda}
\log P(y^*=k|{\bf x}^*) &\approx& \sum_{g=1}^G X_g^* \log d_{kg} - \sum_{g=1}^G s^*\lambda_gd_{kg}  \nonumber \\
                               &&+ \log\pi_k + C,
\end{eqnarray}
where the right hand of (\ref{plda}) is the discriminant score of PLDA.
That is, the NBLDA classifier reduces to the PLDA classifier when there is little dispersion in the data.
From this point of view, the proposed NBLDA can be treated as a generalized version of PLDA.

Since NBLDA contains the dispersion parameter which PLDA does not have,
in what follows, we investigate how the dispersion changes their discriminant scores.
We conduct a numerical comparison between NBLDA and PLDA.
Two cases are considered in this paper.
The first case is that all genes have a common dispersion, and the second is that genes have different dispersions.
Note that the classifiers (\ref{rule.1}) and (\ref{plda}) have the same terms: $\log\pi_k$ and $C$.
Without loss of generality, we compute the discriminant scores only using the first two terms in (\ref{rule.1}) and (\ref{plda}), respectively.
In the comparison study, we fix $X^*_g=10$, $d_{kg}=1.5$, $s^*=1$, $\lambda_g=10$ and $G=500$.
For the case of common dispersion, we set the dispersion ranging from 0 to 20.
For the case of different dispersions, we let $\phi_g$ be independent and identically distributed (i.i.d.) random variables from a chi-squared distribution with the degrees of freedom ranging from 0.1 to 5.

\begin{figure}[!tpb]
\begin{minipage}[t]{0.9\linewidth}
\centering
\scalebox{0.52}{\includegraphics{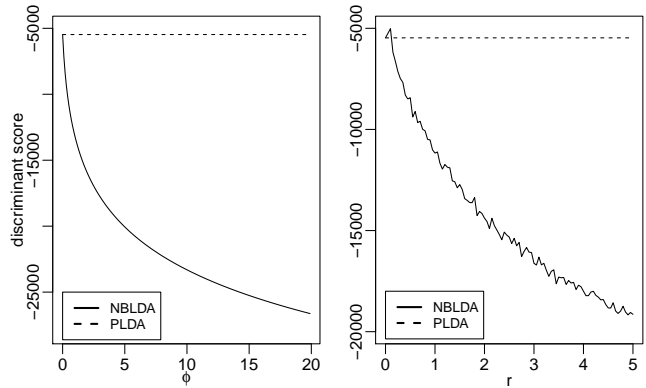}}
\end{minipage}
\caption{Numerical comparisons between NBLDA and PLDA.
The left panel shows the results with a common dispersion $\phi$.
The right panel shows the results with different gene-specific dispersions $\phi_g$ which are i.i.d. random variables from a chi-squared distribution with $r$ degrees of freedom.
We compute the discriminant scores of NBLDA and PLDA for different $\phi$ and $r$.}\label{fig:01}
\end{figure}

Figure 1 exhibits the comparison results.
The left panel shows the results for the case of common dispersion.
Note that the discriminant score of PLDA is independent of the dispersion parameter and hence is a constant.
For NBLDA, its discriminant score is a curve, and the slope is large for low dispersions and small for high dispersions.
We discover that the discriminant score of NBLDA is sensitive to the dispersion.
Even when the dispersion is very small, the difference between the two discriminant scores is significant.
The right panel in Figure 1 shows the results for the case of different dispersions.
The pattern of the right panel is similar to the left one except that the curve of NBLDA is not smooth.
This suggests that when we analyze real data, we should first compute its average dispersion and then use such information to determine which classifier to use.

\subsection{Parameter Estimation}

Note that the discriminant score in (\ref{rule.1}) involves two unknown parameters, size factor ${s}^*$ and dispersion parameter ${\phi}_g$.

\vskip 12pt
\subsubsection{Size factor estimation}

Due to different sequencing depths, samples have different total numbers of reads.
Hence a normalization of the read counts through a size factor is a necessary step for analyzing RNA-Seq data \cite{bullard2010evaluation,dillies2013comprehensive}.
To estimate the size factor ${s}_i$ for the training data and the size factor ${s}^*$ for the test data,
we consider the following three procedures:
\begin{itemize}
\item {\it Total count}:
\citeasnoun{witten2011classification} divided the total read counts of sample $i$ by the total read counts of all samples to estimate the size factor of sample $i$. That is,
$$\hat{s}^*=\frac{\sum_{g=1}^G X^*_g}{\sum_{i=1}^n\sum_{g=1}^G X_{ig}},$$
$$\hat{s}_i=\frac{\sum_{g=1}^G X_{ig}}{\sum_{i=1}^n\sum_{g=1}^G X_{ig}}.$$

\item {\it DESeq}:
\citeasnoun{anders2010differential} first divided the read counts of sample $i$ by the geometric mean of all samples' read counts,
and then estimated the size factor by computing the median of those $G$ values.
Specifically, the size factors are estimated by
$$\hat{s}^*={\rm median}_g\frac{X_g^*}{(\prod_{i=1}^n X_{ig})^{1/n}},$$
$$\hat{s}_i={\rm median}_g\frac{X_{ig}}{(\prod_{l=1}^n X_{lg})^{1/n}}.$$

\item {\it Upper quartile}:
\citeasnoun{bullard2010evaluation} proposed a robust method that uses the upper quartile of the read counts to estimate the size factors.
Specifically, the size factors are estimated by
$$\hat{s}^*=\frac{q^*}{\sum_{i=1}^n q_i},$$
$$\hat{s}_i=\frac{q_i}{\sum_{i=1}^n q_i},$$
where $q^*$ and $q_i$ are the upper quartiles for the test data and sample $i$ in the training data, respectively.
\end{itemize}
In our simulation studies, we find that the three methods provide little difference in the performance of classification.
Hence, for brevity, we only report in the remainder of the paper the simulation results based on the total count method.



\vskip 12pt
\subsubsection{Dispersion parameter estimation}

Various methods for estimating the dispersion parameter $\phi_g$ have been proposed in the literature
\cite{robinson2008small,robinson2010edger,anders2010differential,hardcastle2010bayseq}.
A comparative study is also available in \citeasnoun{landau2013dispersion}
where the authors investigated the influence of different dispersion parameter estimates on detecting differentially expressed genes in RNA-Seq data.
More recently, \citeasnoun{yu2013shrinkage} proposed a shrinkage estimator for $\phi_g$ that shrinks the gene-specific estimation towards a target value.
Specifically, the dispersion estimator is estimated by
\begin{eqnarray}\label{disp}
\hat{\phi}_g = \delta\xi + (1-\delta)\tilde{\phi}_g,
\end{eqnarray}
where $\delta$ is a weight defined as
\begin{eqnarray}
\delta =
\frac{\sum_{g=1}^G\left\{\tilde{\phi}_g-(1/G)\sum_{g=1}^G\tilde{\phi}_g\right\}^2/(G-1)}{\sum_{g=1}^G\left(\tilde{\phi}_g-\xi\right)^2/(G-2)},
\nonumber
\end{eqnarray}
$\tilde{\phi}_g$ are the initial dispersion estimates obtained by the method of moments,
and $\xi$ is the target value calculated by minimizing the average squared difference between $\tilde{\phi}_g$ and $\hat{\phi}_g$.
In this paper, we use the estimator (\ref{disp}) to estimate the dispersion parameter.


\section{Simulation Studies}

In this section, we evaluate and compare the following classification methods:
\begin{itemize}

\item NBLDA,

\item PLDA,

\item Support vector machines (SVM),

\item K-nearest neighbors (KNN).

\end{itemize}
For PLDA, we use the R package ``PoiClaClu" provided in \citeasnoun{witten2011classification}.
For SVM, we use the R package ``e1071" and choose the radial basis kernel in our simulation studies.
For KNN, we choose $k=1$, 3 and 5.

\subsection{Simulation Design}

We generate the data from the following negative binomial distribution:
\begin{eqnarray}
\left(X_{ig}|y_i=k\right) ~\sim~ {\rm NB}(s_i \lambda_g d_{kg},\phi).
\end{eqnarray}
The total number of classes is $K=2$, and both the training data and test data have $n$ samples.
In all $G$ genes, the proportions of differentially expressed genes are 0.2, 0.4, 0.6, 0.8 and 1.0,
which represents that 20\%, 40\%, 60\%, 80\% and 100\% genes are differentially expressed, respectively.
For the differentially expressed genes, we set $\log d_{kg}=z_{kg}$,
where $z_{kg}$ are i.i.d. random variables from the normal distribution $N(0,\sigma^2)$.
For the constant genes, we set $d_{kg}=1$.
The size factors $s_i$ are i.i.d. random variables from the uniform distribution on [0.2, 2.2].
The $\lambda_g$ values are i.i.d. random variables from the exponential distribution with rate 0.04.
Note that, for the sake of fairness, we have essentially followed the same simulation settings as those in \citeasnoun{witten2011classification}.
For the values of $G$, $n$, $\phi$ and $\sigma$, we specify them in Figures 2, 3 and 4.

To compare these methods, we compute the mean misclassification rates as follows:
for each simulation, we generate $n$ test samples and compute the following misclassification rate:
$$\frac{{\rm the~number~of~misclassified~samples}}{n}.$$
We run 1,000 simulations, compute its mean, and then obtain the mean misclassification rate.


\begin{figure}[!tpb]
\begin{minipage}[t]{0.9\linewidth}
\centering
\scalebox{0.52}{\includegraphics{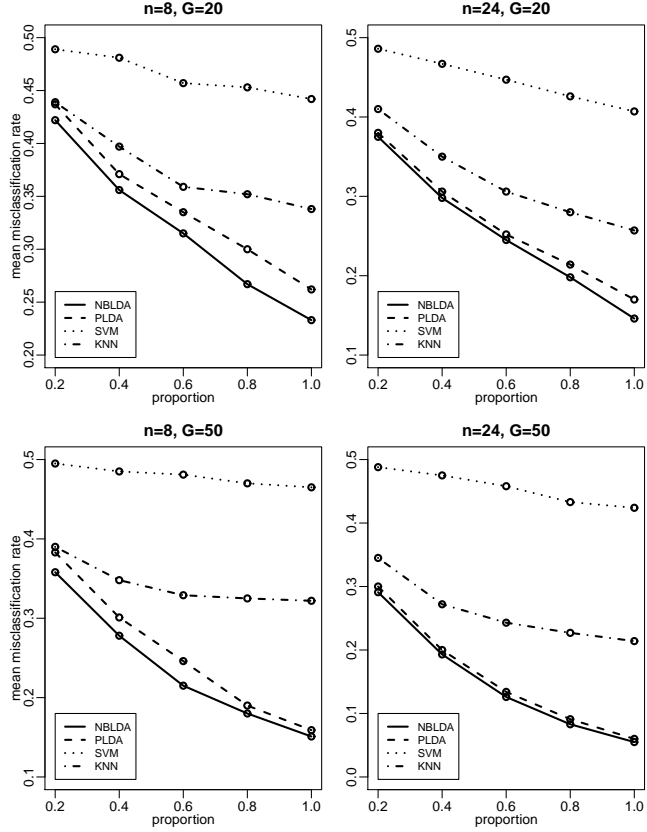}}
\end{minipage}
\caption{Mean misclassification rates for all four methods with $\phi=20$ and $\sigma=5$.
The x-axis represents the proportion of differentially expressed genes.
20\%, 40\%, 60\%, 80\% and 100\% differentially expressed genes are considered, respectively.
These plots investigate the effect of proportion of differentially expressed genes.}\label{fig:01}
\end{figure}

\begin{figure}[!tpb]
\begin{minipage}[t]{0.9\linewidth}
\centering
\scalebox{0.52}{\includegraphics{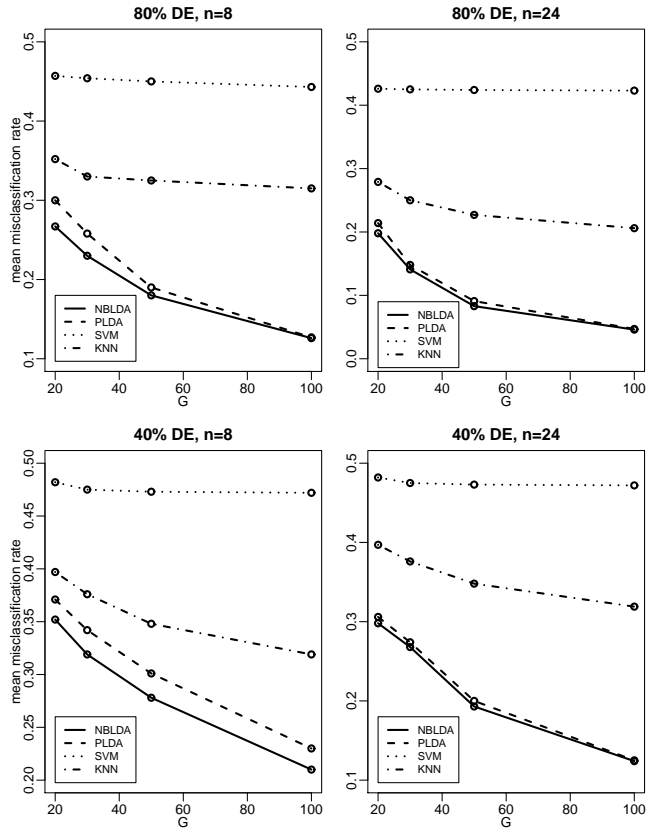}}
\end{minipage}
\caption{Mean misclassification rates for all four methods with $\phi=20$ and $\sigma=5$.
``80\% DE" means 80\% genes are differentially expressed, and the same to ``40\% DE".
This plot investigates the effect of numbers of genes.}\label{fig:02}
\end{figure}

\begin{figure}[!tpb]
\begin{minipage}[t]{0.9\linewidth}
\centering
\scalebox{0.52}{\includegraphics{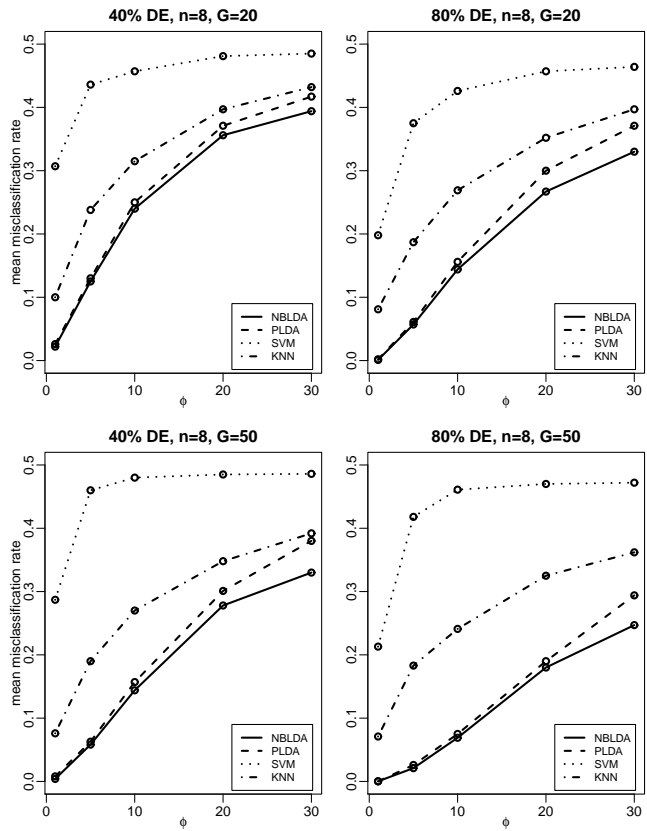}}
\end{minipage}
\caption{Mean misclassification rates for all four methods with $\sigma=5$.
``80\% DE" means 80\% genes are differentially expressed, and the same to ``40\% DE".
This plot investigates the effect of overdispersion.}\label{fig:03}
\end{figure}

\subsection{Simulation Results}

Figure 2 investigates the effect of the proportion of differentially expressed genes on the mean misclassification rate.
In general, with an increasing number of differentially expressed genes, both methods have decreased mean classification rates.
NBLDA always outperforms the other three methods.
In particular, when the sample size is small ($n=8$), NBLDA has a significant improvement over the other approaches.

Figure 3 investigates the impact of the number of genes on the mean misclassification rate.
We consider $G=20$, 30, 50, and 100 for this investigation.
From Figure 3, we observe that an increasing number of genes will lead to a lower misclassification rate.
NBLDA shows its superiority over the other three methods,
and the improvement is more significant when the sample size and the number of genes are smaller.

Figure 4 investigates the effect of overdispersion on the mean misclassification rate.
We consider $\phi=1$, 5, 10, 20 and 30 for this investigation.
Figure 4 shows that a larger dispersion will result in a higher mean misclassification rate.
Both NBLDA and PLDA perform better than SVM and KNN.
When the overdispersion is not very high, NBLDA and PLDA have similar performance, with NBLDA slightly better than PLDA.
When the overdispersion is high, however, the performance of NBLDA is much better than PLDA.

For real biomedical research in which RNA-Seq technology is used,
it is common that thousands or tens of thousands of genes are measured simultaneously.
We perform a gene selection procedure to screen the informative genes before applying a classification rule to RNA-Seq data.
By doing gene selection, we rule out the noise as much as possible so that the variance of the discriminant score is reduced,
and consequently we have an increased interpretability.
For more details, see Section 4.

\section{Real Data Analysis}

We first describe four data sets.
The first three are RNA-Seq data and the last one is a chromatin immunoprecipitation (ChIP) sequencing data set.

\begin{itemize}

\item \textit{Liver and kidney data} \citeasnoun{marioni2008rna}.
There are two classes in this data set. One class contains 7 technical replicates which come from a liver sample.
The other class contains 7 technical replicates which come from a kidney sample.
A total of 22,925 genes are measured in this data set.
The data set is available as a Supplementary File in \citeasnoun{marioni2008rna}.

\item \textit{Yeast data} \citeasnoun{nagalakshmi2008transcriptional}.
The data set contains two library preparations: random haxamer (RH) and oligo (dT), which are treated as two classes in this paper.
In each class, three samples are included: one original sample, its technical replicate, and its biological replicate.
A total of 6,874 genes are quantified in this data set.
The data set is available as a Supplementary File in \citeasnoun{anders2010differential}.

\item \textit{Cervical cancer data} \citeasnoun{witten2010ultra}.
Two groups of samples are contained in this data set. One is the nontumor group which includes 29 samples, and the other one is the tumor group which includes 29 samples.
There are 714 microRNAs in this data set.
This data set is available in Gene Expression Omnibus (GEO) Datasets with access number GSE20592.

\item \textit{Transcription factor binding data} \citeasnoun{kasowski2010variation}.
This data set contains 10 classes with a total of 39 samples.
19,061 binding regions are included in this data set and those regions are treated as distinct features.
This data set is available as a Supplementary File in \citeasnoun{anders2010differential}.

\end{itemize}


\subsection{Gene Selection}

The BSS/WSS method, which is proposed by \citeasnoun{dudoit2002comparison},
is a common gene selection method and has been widely used in the literature
\cite{lee2005extensive,pang2009shrinkage,huang2010bias}.
This method computes the ratio of the sum of squares between groups to the sum of squares within groups for each gene,
and selects genes whose ratios are in the top.
However, this method assumes the data to be normally distributed so that it may not be suitable for RNA-Seq data.

\citeasnoun{witten2011classification} proposed a screening method to select genes for RNA-Seq data.
Since gene $g$ will be deleted from the classification rule, $d_{kg}=1$,
they shrink the estimate of $d_{kg}$ towards 1 by using soft-thresholding to perform the gene selection procedure.
However, this method can not be applied to our discriminant analysis because the dispersion is involved in our discriminant rule.
For the negative binomial distribution,
edgeR \cite{robinson2008small,robinson2010edger} has been proposed to detect differentially expressed genes in RNA-Seq data.
This method first estimates the gene-wise dispersions by maximizing the combination of gene-specific conditional likelihood and common conditional likelihood,
and then replaces the hypergeometric distribution in Fisher's exact test by the negative binomial distribution to construct an exact test.
In this paper, we use edgeR to perform the gene selection procedure, which is available in Bioconductor (www.bioconductor.org).

\begin{figure}[!tpb]
\begin{minipage}[t]{0.9\linewidth}
\centering
\scalebox{0.52}{\includegraphics{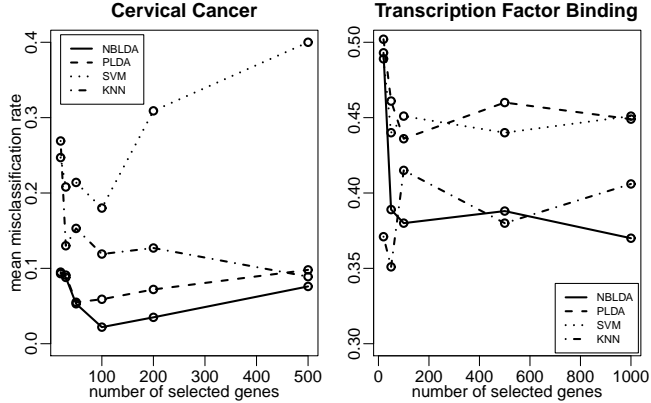}}
\end{minipage}
\caption{Mean misclassification rates for Cervical cancer data
and Transcription factor binding data.}\label{fig:04}
\end{figure}


\subsection{Results}

We first conduct the gene selection procedure using edgeR and obtain $G$ genes for further analysis.
We then randomly split the sample into two sets: the training set and the test set.
The training set is used to construct the classifier and the test set is used to compute the misclassification rate.
We repeat the whole procedure 1,000 times and compute the mean misclassification rate for the four methods, NBLDA, PLDA, SVM, and KNN, respectively.

The comparison results are shown in Figure 5.
Because the mean misclassification rates of the four methods are all zeros for Liver and kidney data and Yeast data,
we only show the results for other two data sets in Figure 5.
For Cervical cancer data, 52 samples are assigned to the training set and 6 samples to the test set.
A total of 20, 30, 50, 100, 200 and 500 genes are selected, respectively.
Among all approaches we consider in this paper, our proposed NBLDA has the lowest misclassification rate.
A big improvement over the other approaches can be observed when more than 50 genes are selected.
For Transcription factor binding data, to conduct the binary classification,
we randomly assign 30 samples to the training set and the remaining 9 samples to the test set.
We choose 20, 50, 100, 500 and 1,000 genes, respectively for this data set.
In Figure 5, we observe that NBLDA also outperforms PLDA for Transcription factor binding data.
Expect when the number of genes is small, NBLDA has a better or comparable performance than the other three methods.

Finally, we estimate the average dispersion of the two data sets to check if it also supports our comparison results made in the previous paragraph.
The simplest way for estimating the dispersion is to use the method of moments.
However, this estimate may not be reliable (sometimes is a negative value) when the sample size is small.
\citeasnoun{landau2013dispersion} and \citeasnoun{yu2013shrinkage} recently reviewed several dispersion estimation methods.
For Cervical cancer data and Transcription factor binding data,
we compute their average dispersions using the method in \citeasnoun{yu2013shrinkage} and present the estimates in Table 1.
We note that both data sets possess a considerably high average dispersion when the number of selected genes is not very large.
This, together with the numerical comparison in Figure 1, explains why NBLDA provides a better performance than PLDA for these two data sets.

\begin{table}[!t]
\caption{The average dispersions for Cervical cancer data and Transcription factor binding data,
where "$G$" represents the number of top genes selected by edgeR. \label{Tab:01}}
{\begin{tabular}{lllll}\hline
Data sets & $G$=20 & $G$=50 & $G$=100 & $G$=500\\\hline
Cervical cancer & 25.71 & 24.42 & 19.02 & 11.03 \\
Transcription factor binding & 8.12 & 5.71 & 4.48 & 2.86\\\hline
\end{tabular}}{}
\end{table}

\section{Discussion}

Next generation sequencing technology has been widely applied in biomedical research and RNA-Seq begins to replace the microarray technology gradually in recent years.
Since RNA-Seq data are nonnegative integers, differing from that of microarray data, it is necessary to develop methods that are well suited for RNA-Seq data.
Two discrete distributions, the Poisson distribution and negative binomial distribution, are commonly used in the literature to model RNA-Seq data.
Compared to the Poisson distribution, the negative binomial distribution allows its variance to exceed its mean and is more suitable for the situations when biological replicates are available.
Nevertheless, the negative binomial model is more complicated than the Poisson model as the additional dispersion parameter also needs to be estimated.

In this paper, we have proposed an NBLDA classifier using the negative binomial model.
Our simulation results show that our proposed NBLDA has a better performance than PLDA in the presence of moderate or high dispersions.
When there is little dispersion in the data, NBLDA is also comparable to PLDA.
We have further explored the relationship between NBLDA and PLDA, and investigated the impact of dispersion on the discriminant score of NBLDA by conducting a numerical comparison.
It is worth noting that even for a small dispersion, the two discriminant scores can be rather different.
This suggests that for real RNA-Seq data with moderate or high dispersion, NBLDA may be a more appropriate method than PLDA.
Note that the true dispersions are unlikely to be known in practice.
Therefore, we propose to first estimate the average dispersion using some novel estimation methods in the recent literature.
Second, if the estimated average dispersion is small,
we use PLDA; and otherwise we use NBLDA.

We note that the independence assumption in \citeasnoun{witten2011classification} and in this paper is very restrictive.
For real gene expression data sets, it may not be realistic to assume that all genes are independent of each other.
In our future study, we would like to incorporate the network information of pathways or gene sets to further improve the performance of classification.
The clustering of sequencing data is also an important issue in biomedical research.
Hence, another possible future work is to extend the clustering method in \citeasnoun{witten2011classification} to follow the negative binomial model.
To conclude, our proposed method is general and can be applied to other next generation sequencing data sets including ChIP-Seq data.

%
%

\section*{Acknowledgements}
Hongyu Zhao's research was supported by the National Institutes of Health grants RC2 DA028909, R01 DA12690, R01 DA12849, R01 DA18432, R01 AA11330, R01AA017535, P50 AA12870, R01 GM59507, R01 DA030976, and UL1 RR024139.
Xiang Wan's research was supported by the Hong Kong RGC grant HKBU12202114, the Hong Kong Baptist University grant FRG2/13-14/005,
and Hong Kong Baptist University Strategic Development Fund.
Tiejun Tong's research was supported by the Hong Kong RGC grant HKBU202711 and the Hong Kong Baptist University grants FRG2/11-12/110, FRG1/13-14/018, and FRG2/13-14/062.


\bibliographystyle{dcu}
\citationstyle{dcu}

%
%
%
%
%
%
%
%

\end{document}